\documentstyle[12pt,epsfig]{article}
\def\b{\bar}
\def\d{\partial}

\def\cA{{\cal A}}

\def\m{\mu}
\def\n{\nu}

\def\t{\tau}

\def\~{\tilde}
\def\h{\eta}

\def\bY3{\bar Y_{,3}}
\def\Y3{Y_{,3}}
\def\z{\zeta}
\def\Z{{\b\zeta}}
\def\Y{{\bar Y}}

\def\`{\dot}
\def\be{\begin{equation}}
\def\ee{\end{equation}}
\def\bea{\begin{eqnarray}}
\def\eea{\end{eqnarray}}

\def\fn{\footnote}

\def\cF{{\cal F}}

\def\mn{{\mu\nu}}

\begin{document}

\title{Twistorial Analyticity and Three Stringy Systems of the Kerr Spinning
Particle}

\author{Alexander Burinskii\\
Gravity Research Group, NSI Russian
Academy of Sciences\\
B. Tulskaya 52, 115191 Moscow, Russia}
\maketitle

\begin{abstract}
\noindent
The Kerr spinning particle has a remarkable analytical twistorial structure.
Analyzing  electromagnetic excitations
of the Kerr circular string which are aligned to this structure, we obtain
 a simple stringy skeleton of the spinning particle which is formed by a
topological coupling of the Kerr circular singular string and by an axial
singular stringy system.

 We show that the chiral traveling waves, related to an orientifold
world-sheet of the axial stringy system, are described
by the massive Dirac equation, so we argue that the axial string may
play the part of a stringy carrier of wave function and play also a dominant
role in the scattering processes.

A key role of the third, {\it complex} Kerr string  is discussed.
We conjecture that it may be one more alternative to the Witten twistor
string, and a relation to the spinor helicity formalism is also discussed.
\end{abstract}

\section{Introduction and basic results}

The Kerr geometry  has  found application in a very wide range
of physical systems: from the rotating black holes and galactic nucleus
to  fundamental solutions of the low energy string theory. It displays
also some remarkable relations to the structure of spinning particles
\cite{Car,Isr,Bur0,IvBur,Lop,BurSen,BurStr,BurSup,BurBag,New}.
Angular momentum of the spinning particles is very high and
the black-hole horizons disappear. Image of the solution changes drastically
taking the form of a naked ringlike source.
In our old paper \cite{Bur0}
this ring was considered as a gravitational
waveguide  caring the traveling electromagnetic waves which
generate the spin and mass of the Kerr spinning particle forming a
microgeon with spin.
It was conjectured \cite{IvBur} that the Kerr ring
represents a closed string, and the traveling
waves are the string excitations. It was noted in \cite{BurSen}
that in the axidilatonic version
of the Kerr solution the field around this ring is similar to the field
around a heterotic string, and recently, it was shown that the Kerr ring is a chiral D-string having
an orientifold world-sheet \cite{BurOri}.

We show here that an analytical twistorial structure
of the Kerr spinning particle leads to the appearance of an
extra axial stringy system.
As a result, the Kerr spinning particle acquires a simple stringy
skeleton \cite{BurTwo,BurAxi} which is formed by a topological coupling
of the Kerr circular string and the axial stringy system, see fig. 1.
Meanwhile, there is a third, {\it complex}
string of the Kerr spinning particle which is also
connected to the Kerr's holomorphic twistorial structure.

In the very interesting recent paper \cite{WitTwi} Witten suggested
a string model with a twistorial target space and argued that some remarkable
analytical properties of the perturbative scattering amplitudes in
Yang-Mills theory \cite{BDK} have the origin in an holomorphic structure of
this string resulting to the holomorphy of the
maximally helicity violating (MHV) amplitudes \cite{Nai,ChaSie}.
However, Witten considered this proposal as tentative \cite{WitTwi},
and other models of strings related to twistors were also suggested.
In particular, the original twistor string suggested about fifteen years ago
by Nair \cite{Nai} was recently modified by  Berkovits \cite{Ber} and
represented as an alternative to the Witten twistor string. Some
other suggestions were also discussed by Siegel \cite{Sie}.

\begin{figure}[ht]
\centerline{\epsfig{figure=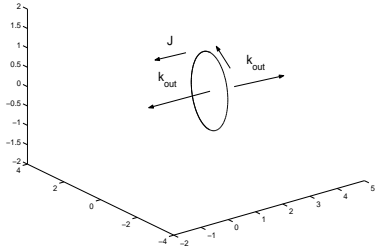,height=4cm,width=5cm}} \caption{Stringy
skeleton of the Kerr spinning particle. Circular D-string and axial stringy
system consisting of two semi-infinite D-strings of opposite chiralities.}
\end{figure}

In this connection we pay the attention to the complex Kerr string
\cite{BurStr} which plays the key role in the stringy systems of the Kerr
geometry and may also represent an alternative to the twistor string.

Contrary to the strings in twistor space, the target space of the complex
Kerr string is ${\bf CM}^4$ which is the base of a twistor line bundle,
so the twistors are adjoined to each point of this string.
In many respects this string is similar to the well known $N=2$ strings.
\fn{Note, that there is a cojecture \cite{NeiVaf} that the Witten
twistor string on ${\bf CP}^{3|4}$ is indeed equivalent to the $N=2$ string.}

The complex string appears naturally in the
initiated by Newman {\it complex representation of the Kerr geometry}
 \cite{BurStr,BurSup,New,BurNst}, where the Kerr source is
generated by a complex world line $X_0^\m (\t)$. Since complex time $\t
=t+i\sigma$ consists of two parameters $t$ and $\sigma$, it parametrizes a
world sheet, and therefore, the complex world lines can be considered as
strings \cite{BurStr,OogVaf}. It turns out that the Kerr complex string
is an open string, and its ends are stuck to the two semi-infinite strings
of the axial stringy system shown in fig.1.
These semi-strings are the lightlike strings of opposite chiralities,
and we show that their interplay is described by the massive Dirac equation.
So, the axial stringy system may be considered as a stringy carrier of the
wavefunction of the Kerr spinning particle.

Therefore, the Kerr spinning
particle displays a very interesting stringy system consisting from the Kerr
circular string, the axial semi-strings and the complex Kerr string.  We show
here that these strings have similar analytical properties, and consistency
of this system is provided by a common orientifold structure.

Returning to the remarkable simplification of the MHV relativistic scattering
amplitudes, one can wonder - which part of the complicated stringy structure
of the Kerr spinning particle may be responsible for the very
simple holomorphy description of the MHV amplitudes.
We discuss it in the end of the paper, arguing that a dominant contribution
to the relativistic scattering process may be coming from
one of the chiral semi-strings of the Kerr axial stringy system.

In our treatment we prefer to work in the Kerr-Schild formalism  \cite{DKS}
which is based on the Kerr theorem which is related to twistors.
 For the reader convenience we describe here briefly the necessary details
of the real and complex structures of the Kerr geometry.
For more details see \cite{BurStr,BurSup,BurNst}.

\section{Twistorial structure of the Kerr congruence}

The Kerr-Schild form of the metric is

\be g_{\m\n}
= \h_{\m\n} + 2 h k_{\m} k_{\n}, \label{ks} \ee
where $ \h_\mn $
is the metric of auxiliary Minkowski space-time  (signature $-+++$) in the
Cartesian coordinates $x^\m=(t,x,y,z)$,
$ h= \frac {mr-e^2/2} {r^2 + a^2 \cos^2 \theta},$ and $k_\m$
is a twisting null field which forms the Kerr principal null congruence (PNC)
 - a family of the  geodesic and shear free null curves.  The Kerr PNC is
shown in Fig. 2. Each null ray of the PNC represents the twistor
$Z^a=\{\m ^\alpha, \omega _{\dot \alpha} \},$ in which spinor $\m ^\alpha$
determines the null direction, $k_{\n} = \bar\m ^{\dot \alpha} \sigma_{\n \dot
\alpha \alpha}\m ^\alpha$, and $\omega _{\dot \alpha}=x^\n \sigma_{\n \dot
\alpha \alpha} \m ^\alpha $ fixes the position (equation) of the ray.

The Kerr's geodesic and shear-free PNC is determined  by {\it the Kerr
theorem} via the solution $Y(x)$ of the algebraic equation $F=0$,
where $F(Y,\lambda _1, \lambda _2)$ is an arbitrary holomorphic function
of the projective twistor coordinates $\{Y,\quad \lambda
_1 = \zeta-Y v,\quad \lambda _2=u+Y\bar\zeta\} = Z^a/Z^0, $ and
\bea 2^{1\over2}\z = x+iy , \quad 2^{1\over2} \Z &=& x-iy , \nonumber \\
2^{1\over2}u = z - t , \quad 2^{1\over2}v &=& z + t  \label{ncc}\eea
are the null Cartesian coordinates.

In the Kerr-Schild formalism the projective spinor field $Y(x)= \m ^2/\m ^1$
determines the null field $k^\m (x)$ and other parameters of the solution.
In particular, the Kerr-Schild null tetrad is given by
\begin{eqnarray}
e^1 &=& d \zeta - Y dv, \qquad  e^2 = d \bar\zeta -  \bar Y dv, \nonumber \\
e^3 &=&du + \bar Y d \zeta + Y d \bar\zeta - Y \bar Y dv, \nonumber\\
e^4 &=&dv + h e^3\label{KSt},
\end{eqnarray}
and the
system of equations $F(Y)=0;\quad dF(Y)/dY =0$ determines
the singular lines which are caustics of the PNC.

In the Kerr solution function $F$ is quadratic in $Y$, and the equation
$F=0$ and the system $F(Y)=dF(Y)/dY =0$ can be explicitly
resolved \cite{BurNst} yielding the structure of the Kerr PNC and the Kerr
singular ring shown in the Fig.2.

\begin{figure}[ht]
\centerline{\epsfig{figure=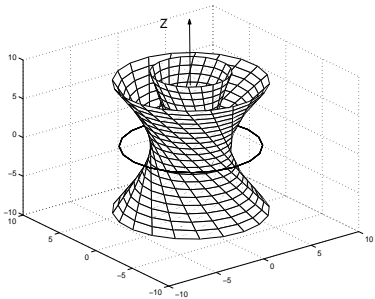,height=4cm,width=5cm}}
\caption{The Kerr circular string and null rays of the PNC
which propagate from `negative' sheet of space-time
onto `positive' one.}
\end{figure}

  The Kerr singular ring is the branch line of the
Kerr space on two sheets: `positive'  ($r>0$) and `negative'
($r<0$) ones, so the Kerr PNC propagates from the `negative' sheet
onto `positive' one through the disk spanned by the Kerr ring.
Since the PNC determines a flow of radiation (in radiative solutions),
one sees that outgoing radiation is compensated by an ingoing flow on the
negative sheet, so the negative sheet acquires the interpretation as a
list of advanced fields which are related to the vacuum zero point fields
\cite{BurOri,BurNst,BurTwo}.
In this interpretation, the wave excitations can be treated as a result of
resonance of a zero point field on the Kerr ring, in the spirit of the
semiclassical Casimir effect. It is remarkable, that similar to the
quantum case, such excitations of the Kerr string take place without
damping since outgoing radiation is compensated by ingoing one.
 Due to the twofolded topology,
the Kerr circular ring turns out to be the ``Alice'' string,
which corresponds to a very minimal nonabelian generalization of the
Einstein-Maxwell system. A truncation of the negative
sheet leads to the appearance of a source in the form of
a relativistically rotating disk \cite{Isr} and to the class of
the superconducting disklike \cite{Lop} and baglike \cite{BurBag}
models of spinning particle.

\section{Axial stringy system}

Let us consider solutions for traveling waves -
electromagnetic excitations of the Kerr circular string.
The problem of electromagnetic excitations of the Kerr black hole
has been intensively studied
as a problem of the quasinormal modes. However, compatibility with the
holomorphic structure of the Kerr space-time put  an extra
demand on the solutions to be aligned to the Kerr PNC, which takes the form
$F^\mn k_\m=0$.
 The aligned wave solutions for electromagnetic fields
on the Kerr-Schild background were obtained in the Kerr-Schild
formalism \cite{DKS}. We describe here only the result referring
for details to the papers \cite{BurTwo,BurAxi}. Similar to the
stationary case \cite{DKS} the general aligned solution is
described by two self-dual tetrad components $\cF _{12} = AZ^2$
and $\cF _{31} = \gamma Z -(AZ),_1$, where function $A$ has the
form \be A= \psi(Y,\t)/P^2, \ee
 $P=2^{-1/2}(1+Y\Y)$, and $\psi$ is an arbitrary holomorphic function of
$\t$ which is a complex retarded-time parameter. Function
$Y(x) = e^{i\phi} \tan \frac {\theta} 2 $ is a projection
of sphere on a complex plane. It is singular
at $\theta=\pi$, and one sees that such a singularity will be present in
any holomorphic function $\psi (Y)$. Therefore, {\it all the aligned e.m.
solutions turn out to be singular at some angular direction $\theta$}.
 The simplest modes
\be
\psi _n = q Y^n \exp {i\omega _n \t}
\equiv q (\tan \frac \theta 2)^n \exp {i(n\phi + \omega _n \t)}
\label{psin}
\ee
can be numbered by index
$n=\pm 1, \pm 2, ...$,
which corresponds to the number of the
wave lengths along the Kerr ring. The mode $n=0$ is the
Kerr-Newman stationary field.
Near the positive $z^+$ semi-axis we have $Y\to 0$  and
near the negative $z^-$ semi-axis  $Y\to \infty$.

Omitting the longitudinal components and the radiation field $\gamma$
one can obtain \cite{BurTwo,BurAxi} the form of the leading wave terms
\be \cF |_{wave} =f_R \ d \z \wedge d u  +
f_L \ d \Z \wedge d v ,
\label{cFLR}
\ee
where
$f_R = (AZ),_1 ; \qquad f_L =2Y \psi (Z/P)^2 + Y^2 (AZ),_1$
are the factors describing the ``left'' and ``right"  waves propagating
along the $z^-$ and $z^+$ semi-axis correspondingly.

The behavior of function $Z=P/(r+ia \cos \theta)$ determines a
singularity of the waves at the Kerr ring,
so the singular waves along the ring induce, via function $Y$,
singularities at the $z^\pm$ semi-axis. We are interested in the asymptotical
properties of these singularities.  Near the $z^+$ axis  $|Y|\to 0$,
and  by $r \to \infty $, we have $Y \simeq  e^{i\phi} \frac \rho {2r}$ where
$\rho$ is the distance from the $z^+$ axis. Similar, near the $z^-$ axis $Y
\simeq  e^{i\phi} \frac {2r} \rho  $ and $|Y|\to \infty$.
The parameter $\t=t  -r -ia \cos \theta$ takes near the z-axis the values $\t _+ = \t |_{z^+}=
t-z-ia,\quad \t _- = \t |_{z^-} =t+z +ia.$

For $|n|>1$ the solutions contain the axial
singularities which do not fall of  asymptotically, but are increasing that
means instability.
Therefore, only the wave solutions with $n=\pm 1$ turn out to be admissible.
The leading singular wave for $n=1$,
\be
\cF^-_1=\frac {4q e^{i2\phi+i\omega _{1} \t_- }} {\rho ^2} \ d \Z \wedge dv ,
\ee
propagates to $z=-\infty$ along the $z^-$ semi-axis.

The leading wave for $n=-1$,
\be
\cF^+_{-1}= -
\frac {4q e^{-i2\phi+i\omega _{-1} \t_+ }} {\rho ^2} \ d \z \wedge du ,
\ee
is singular at $z^+$ semi-axis and propagates to $z=+\infty$.
The described singular waves can also be
obtained from the potential
$\cA^\m= - \psi (Y,\t)(Z/P) k^\m $.
The $n=\pm 1$ partial solutions $\cA^{\pm} _n$ represent asymptotically
the singular plane-fronted e.m. waves propagating along
$z^+$ or $z^-$ semi-axis without damping.
The corresponding
self-consistent solution of the Einstein-Maxwell field equations
are described in \cite{BurTwo}. They are singular plane-fronted waves
 having the Kerr-Schild form of metric (\ref{ks}) with a constant vector
$k^\m$.  For example, the wave propagating along the $z^+$ axis has
$k^\m dx^\m= - 2^{1/2}du $).
The Maxwell equations take the form
$\Box \cA = J=0 , $ where $Box$ is a flat D'Alembertian,
and can easily be integrated leading to the solutions
$ \cA ^+ = [ \Phi ^+(\z) + \Phi ^-(\Z) ] f^+(u) du, \qquad
 \cA ^- = [ \Phi ^+(\z) + \Phi ^-(\Z) ] f^-(v) dv,$
where $\Phi ^{\pm}$ are arbitrary analytic functions, and functions
$f^\pm $
describe the arbitrary retarded and advanced waves.
Therefore, the wave excitations of the Kerr ring lead to the appearance
of singular pp-waves which propagate outward along the $z^+$ and/or
$z^-$ semi-axis.

These axial singular strings are evidences
of the axial stringy currents, which are exhibited explicitly
when the singularities are regularized. Generalizing the field model to
the Witten field model for the cosmic superconducting strings \cite{Wit},
one can show \cite{BurAxi}  that these singularities are replaced by the
chiral superconducting strings, formed by a condensate of the Higgs field, so
the resulting currents on the strings are matched with the external gauge
field.

The stringy system containing the chiral modes of only
one direction cannot exist since it is degenerated in a word-line
\cite{BurOri}. A combination of two $n=\pm 1$ excitations
 leads to the appearance of  two
semi-infinite singular D-strings of opposite chirality as it is
shown at the fig.1.
Similar to the all ray of PNC,  the semi-infinite singularities
can be extended to the negative sheet passing through the Kerr ring.
The world-sheet of such a system acquires the structure of an orientifold
and is given by
\be
x^\m (t,z)=\frac 12 [ (t-z)k_R^{\m} + (t+z)k_L^{\m}],
\label{astr}
\ee
where the lightlike vectors $k^\m$ are constant and normalized.
At the rest frame the time-like components are equal
$k_R^{0}  =k_L^{0}=1$, and the space-like components are oppositely directed,
$k_R^{a} + k_L^{a}=0, \quad a=1,2,3.$
Therefore,
$\dot x^\m =(1,0,0,0),$ and $ x'^\m =(0,k^a),$
and the Nambu-Goto string action
$S=\alpha^{\prime -1}\int\int\sqrt{(\dot x)^2 ( x')^2 - (\dot x x')^2 } dtdz
$ can be expressed via $k_R^{\m}$ and $k_L^{\m}$.

For the system of two D-strings  in the rest one can use the gauge with
$\dot x^0 =1, \quad \dot x^a=0,$ where the term   $(\dot x x')^2$ drops out,
and  the action takes the form
$S=\alpha^{\prime -1}\int dt \int \sqrt{p^a p_a} d\sigma,$
where $p^a = \d _\sigma x^a = \frac 12 [ x_R^{\prime\m}(t+\sigma) -
x_L^{\prime\m}(t-\sigma)].$

To normalize the infinite string  we have to perform a renormalization
putting      $ m=\alpha^{\prime -1}\int (x')^2 dz , $
which yields the usual action for the center of mass of a pointlike particle
$ S= m\int \sqrt{( \dot x)^2 }dt.$

\section{The Dirac equation}

It is known that in the Weyl basis the Dirac current can be
represented as a sum of two lightlike components of opposite
chirality
\be
J_\m = e (\bar \Psi \gamma _\m \Psi) = e (\chi ^{+} \sigma _\m  \chi +
\phi ^{+} \bar \sigma ^\m  \phi ),
\ee
where
$\Psi =
\left(\begin{array}{c}
 \phi _\alpha \\
\chi ^{\dot \alpha}
\end{array} \right).$
It allows us to conjecture that the Dirac equation may describe  the
Kerr axial stringy system - the lightlike currents of two opposite
chiralities which are positioned on the different folds of the world-sheet.

In this case the four-component spinor $\Psi$, satisfying the massive Dirac
equation
$(\gamma^\m \hat P_\m -m)\Psi=0,  \qquad
\hat P_\m =i \hbar \d _\m, $
describes an interplay of the axial string currents on the different folds
of the Kerr space.
In the Weyl basis the Dirac system splits into
\be
m \phi _\alpha  =
i \sigma ^\m _{\alpha \dot \alpha} \d_\m \chi ^{\dot \alpha}, \qquad
m \chi ^{\dot \alpha}  =
i \bar\sigma ^{\m \dot\alpha \alpha}  \d_\m \phi _{\alpha}.
\label{DirSpl}
\ee
Regarding the function $Y$ as a projective spinor field
$Y=\phi _2 /\phi _1 , $ one sees that
near the $z^+$ semi-axis $Y\to 0$, so one can set in this limit
${\phi}_{\alpha} =
\left(\begin{array}{c}
 {\phi}_{1}\\
\phi _{2}
\end{array} \right) =
\left(\begin{array}{c}
 1\\
0
\end{array} \right).$
This spinor describes the lightlike vector
\be
k_{R} =d(t-z)=
\bar  \phi _{\dot\alpha}
\bar\sigma _\m^{\dot\alpha\alpha} dx^\m  \phi _\alpha .
\label{kR}
\ee
Similar, near  the $z^-$ semi-axis $\bar Y \to \infty$,
and this limit corresponds to the spinor
${\bar \chi}^{\alpha} =
\left(\begin{array}{c}
 1\\
0
\end{array} \right) $
which describes the lightlike vector
\be
k_{L} =d(t+z)=
\bar \chi ^{\alpha}
\sigma _{\m\alpha\dot\alpha} dx^\m  \chi ^{\dot\alpha} .
\label{kL}
\ee
The vectors $k_L$ and $k_R$ are the generators of the
left and right chiral waves correspondingly.
The functions $\phi$ and $\chi$ are fixed up to arbitrary
gauge factors. Forming the four component spinor function
\begin{equation}
\Psi = {\cal  M}(p_\m x^\m)
\left(\begin{array}{c}
a \phi _\alpha \\
b \chi ^{\dot \alpha}
\end{array} \right),
\label{MPsi}\end{equation}
we obtain  from the Dirac equations (\ref{DirSpl})

$am=(p_0 -p_z) b \ln {\cal  M}' , \qquad
bm=(p_0 +p_z) a \ln {\cal  M}',$
$p_x +ip_y =0,$
which realizes the Dirac idea on splitting the
relation   $p_0^2 = m^2 +p_z^2, $
and
 \be {\cal M}=e^{-i \omega t + i z p_z }.\ee
  This solution describes a wavefunction  of a free spinning particle
moving along the z-axis. It oscillates with the
Compton frequency, which is determined by excitations of the Kerr
circular string, and leads to a
 plane fronted modulation of the axial string by de Broglie periodicity.
It allows us to conjecture that the axial stringy system acquires in the
Kerr spinning particle the role
of {\it a stringy carrier of wavefunctions}.

\section{Orientifold and a complex Kerr string}

The tension for a free particle tends to zero, but it can be finite for the
bounded states when the axial string forms the closed loops.
An extra tension can appear in the bounded systems, since an extra magnetic
flow can concentrate on the closed loops.
By formation of the closed loops chiral modes have to be matched forming an
orientifold structure. The position $x^\m$ of the (say)`left'
semi-string, being extended  to the `negative' sheet, has to coincide
with the position of the `right' one. For a free stationary particle it is
realized. However, in general case it puts a very strong restriction on the
dynamics of the string.
The orientifold  of the axial string is also suggested by the structure
of superconducting strings. As it pointed out by Witten \cite{Wit},
it contains the light-like fermions of both the chiralities, moving in
opposite directions, so the massive Dirac solutions for a superconducting
string describe an interplay of the trapped fermions of the opposite
chiralities. In general case orientifolding the world-sheet requires a
very high degree of symmetry for the string.
It turns out that this is provided by the holomorphy and orientifold
structure of the third, {\it complex string} of the Kerr geometry
\cite{BurStr}.

  The complex representation of the Kerr geometry
\cite{BurNst,BurStr,BurSup,New}
is generated by a complex world line $X_0^\m(\t)\in {\bf CM}^4,$ where
$\t=t_0+i\sigma$ is a complex time parameter. In the stationary Kerr case
$X_0^\m(\t)=(\t,0,0,ia).$
The complex world line forms a world sheet and takes an intermediate position
between particle and string \cite{BurStr,OogVaf}.  The real fields on the
real space-time $x^\m$ are determined via a retarded-time construction, where
the vectors $K^\m=x^\m - X_0^\m(\t)$ have to satisfy the complex light-cone
constraints $K_\m K^\m =0$.
Looking on the complex retarded-time equation
\be t-\t \equiv t-t_0 -i\sigma =\tilde r , \ee
where $\tilde r = r+i a \cos \theta$ is the Kerr complex radial distance
($\theta$ is a direction of the null ray (twistor), and $r$ is the
spatial distance along the ray), one sees that on the real space-time
$\sigma= - a \cos \theta .$  Therefore, the light-cone constrains
select a strip on $\t$ plane, $\sigma \in [-a,+a]$, and the
complex world sheet $X_0^\m(t,\sigma)$ acquires the
boundary, forming {\it an open complex string} with target space ${\bf CM}^4$
which is a base of a twistor bundle.  In many aspects this string is similar
to the $N=2$ string \cite{OogVaf,GSW}, but it has the signature $(-+++)$ and
the euclidean world sheet.

The complex light cones, adjoined to each point of the world
sheet, split into the `right' and `left' null planes which
are the twistors having their `origins' $X_0^\n$ at the
points of the world sheet. The `left' null planes - twistors
$Z^a=\{\m ^\alpha, X_0^\n \sigma_{\n \dot \alpha \alpha} \m
^\alpha \}$ - form a holomorphic twistor subspace.  The two twistors
which are joined to the ends of the complex string, $X_0^\n (t\pm
ia)$, have the directions $\theta= 0,\pi$ and are generators of the singular
$z^\pm$ semi-strings, so the complex string turns out to be a D-string which
stuck to two singular semi-strings of opposite chiralities.
Therefore, $z^\pm$ singular strings may carry the Chan-Paton factors which
will play the role of quarks with respect to the complex string.
Extending this analogy one can assume that the Kerr circular
lightlike string, adjoined to the `center' of the complex string $\sigma =0$,
may carry the factors of the third quark of this stringy system.  In
the nonabelian generalizations, the chiral singular strings may carry the
color currents \cite{Nai} connected with color traveling waves \cite{Sie1}
which play the role of the color quarks.

It was obtained in \cite{BurStr,BurOri} that boundary conditions of the
complex string demand the orientifold structure of the world-sheet.
 The resulting field equations have the
solutions which are satisfied by the holomorphic (!) modes
$X_0^\m(\t)$. These solutions are the `left' modes, and the
`right' modes appear by orientifolding. The interval for parameter
$\sigma=a\cos \theta$ is doubled:  $\Sigma _\pm =[-a,a]$, forming a circle
$S^1 = \Sigma _+ \bigcup \Sigma _-$ which is parametrized by
$\theta \in [0,2\pi]$, so
  $\Sigma _-$ parametrizes the string in opposite orientation.
 Orientifold is formed by a $Z_2$ factorization of the world-sheet.
The string turns into a closed but folded one.
The `right' modes have the form of the `left' ones but have
a support on the reversed interval
$\theta \in [\pi,2\pi] \in\Sigma _-$:

\be X_{0R}^\m (t+ia\cos(2\pi - \theta)) = X_{0L}^\m (t+ia\cos\theta), \ee
which is a well-known kind of extrapolation \cite{GSW}.
So, for the period $(0,2\pi)$ occurs a flip of the modes $left \to right $.
It is accompanied by
a space reversal of the twistors joined to the points of the world-sheet,
which is performed by the exchange $Y \to - 1/\bar Y$.
As a result the orientifold structure of the complex string
provides the orientifold structure for the chiral axial strings.

It can be shown that all the three orientifold structures
can be joined forming a unite orientifold matched to the holomorphic
twistorial structure of the Kerr space-time. We intend to discuss it
in details elsewhere.
\section{Relation to the spinor helicity formaslism}

Let us finally discuss the possible relation to the MHV scattering
amplitudes . For the massless and
relativistic particles with spin, the external line factors of the
amplitudes are described in the spinor helicity formalism
\cite{WitTwi,BDK,Nai,ChaSie} which is
based on a color decomposition and a reduced description in terms of
the lightlike momentum
$p_{\alpha \dot \alpha} = \sigma^\m_{\alpha \dot \alpha}p_\m $
and $\pm$ helicities.

The momentum $p_{\alpha \dot \alpha}$ is represented via
the undoted spinors
$<p| =p^\alpha $, $\quad |p> =p_\alpha $ and doted spinors
$[p| =p^{\dot\alpha} $, $\quad |p] =p_{\dot\alpha}$
in the form

\be p=\pm|p>[p| . \ee

Since the Kerr spinning particle is massive, its momentum along
the z-axis  is represented as a sum of the lightlike parts $p^\m=
p_L^{\m} + p_R^{\m}$, where the corresponding spinors are

\be p_L^\alpha=<p_L|=p_L\cdot <k_L| \ee

and

\be p_R^\alpha=<p_R|= p_R \cdot<k_R| .\ee

For a relativistic motion we have either $p_L << p_R$ or $p_L>>p_R$,
which determines the sign of helicity, and as a result one of the
axial semi-strings turns out to be strongly dominant for the scattering.
  It may justify the use of the very reduced description of the Kerr
spinning particle via the spinor helicity formalism.
We are led to the conclusion that the axial
stringy system may be responsible for the high energy scattering
processes.

\section*{Acknowledgments}
It is a pleasure to thank Sergi Hildebrandt for reading the text and
useful remarks, and also Valeri Frolov for useful conversations.
This work is supported by the RFBR grant 04-0217015-a
and by the research grant of the ISEP foundation by Jack Sarfatti.

\end{document}